\documentclass[12pt]{article}
\usepackage{graphicx}

\newcommand\pubnumber{}
\newcommand\pubdate{\today}

\def\narawu{Department of Physics\\
Nara Women's University, Kita-Uoya-Nishi-machi 630-8506 Nara, JAPAN}
\def\support{\footnote{Author's participation was supported by 
MEXT KAKENHI, Grant-in-Aid for Scientific Research on Innovative Areas, 
entitled ``Elucidation of New hadrons with a Variety of Flavors''.}}

\def\Title#1{\begin{center} {\Large #1 } \end{center}}
\def\Author#1{\begin{center}{ \sc #1} \end{center}}
\def\Address#1{\begin{center}{ \it #1} \end{center}}

\newcommand\pubblock{\rightline{\begin{tabular}{l} \pubnumber\\
         \pubdate  \end{tabular}}}
\newenvironment{Abstract}{\begin{quotation}  }{\end{quotation}}
\newenvironment{Presented}{\begin{quotation} \begin{center} 
             PRESENTED AT\end{center}\bigskip 
      \begin{center}\begin{large}}{\end{large}\end{center} \end{quotation}}
\def\Acknowledgements{\bigskip  \bigskip \begin{center} \begin{large}
             \bf ACKNOWLEDGEMENTS \end{large}\end{center}}

\def\beq{\begin{equation}}
\def\eeq#1{\label{#1}\end{equation}}
\def\eeqn{\end{equation}}

\def\beqa{\begin{eqnarray}}
\def\eeqa#1{\label{#1}\end{eqnarray}}
\def\eeqan{\end{eqnarray}}

\let\bar=\overbar

\def\Dslash{\not{\hbox{\kern-4pt $D$}}}
\def\dslash{\not{\hbox{\kern-2pt $\del$}}}

\def\msb{{\bar{\ssstyle M \kern -1pt S}}}

\begin{document}
\begin{titlepage}
\pubblock

\vfill
\Title{Determination of $\phi_1^{\rm eff} (\beta_{\rm eff})$ by 
$B^0 \rightarrow K^+ K^- K^0_S$, $\pi^+ \pi^- K^0_S$ and $K^0_S K^0_S K^0_S$
decays}
\vfill
\Author{ Kenkichi Miyabayashi \support}
\Address{\narawu}
\vfill
\begin{Abstract}
Measurements of the $CP$ violation parameter 
$\phi_1^{\rm eff} (\beta_{\rm eff})$ by $b \rightarrow s$
penguin mediated three-body $B$ decays, $B^0 \rightarrow K^+ K^- K^0_S$,
$\pi^+ \pi^- K^0_S$ and $K^0_S K^0_S K^0_S$ are reviewed in this report.
\end{Abstract}
\vfill
\begin{Presented}
6th International Workshop on the CKM Unitary Triangle (CKM2010)\\
University of Warwick, United Kingdom,  September 6-10, 2010
\end{Presented}
\vfill
\end{titlepage}
\def\thefootnote{\fnsymbol{footnote}}
\setcounter{footnote}{0}

\section{Introduction}

Time-dependent $CP$ violation measurements started with 
tree-mediated $B$ decays such as $B^0 \rightarrow J/\psi K^0$
in order to perform critical tests of Kobayashi-Maskawa theory.
The most promising measurement is the $CP$ violation caused by
interference between mixing and decay. If we choose the 
$B$ meson decay mode caused by the amplitude containing no complex phase, 
$CP$ violation arises from the $B^0 - \overline{B}^0$ mixing
that contains one of the Kobayashi-Maskawa matrix element $V_{td}$
and thus is directly related to the $CP$ violation angle $\phi_1 (\beta)$
\footnote{$\phi_1$ is used hereafter.}.

Penguin-mediated $B$ decays are sensitive to New Physics, 
which can provide observed $CP$-violating parameters different from
those obtained in tree-mediated processes.
In the Standard Model (SM), the $b \rightarrow s$ transition
does not contain a complex phase in the decay amplitude and thus is 
expected to exhibit the same time-dependent $CP$ violation as 
$B^0 \rightarrow J/\psi K^0$. 
In other words, the New Physics effect contribute to the 
decay amplitude because of its one-loop nature and could appear 
as possible deviations of $CP$ violation parameters from 
the SM expectation. 
The first round of measurements used a quasi-two-body approach,
i.e. $B^0 \rightarrow \phi K^0$, $\rho^0 K^0_S$, and so on.
However, several contributions are overlapping because of relatively 
wide natural widths of the involved resonances.
For example, in $B^0 \rightarrow K^+ K^- K^0_S$ case, 
$\phi$, $f_0$ and other resonant or non-resonant 
contributions are there, and they interfere with each other.
Therefore, in order to resolve those interfering contributions 
into three-body final state such as $K^+K^-K^0_S$ or 
$\pi^+\pi^- K^0_S$, the time-dependent Dalitz distribution is fitted 
to extract $CP$ violation parameters.
By this technique, we can measure $\phi_1^{\rm eff}$ and ${\cal{A}}_{CP}$
which denote mixing-induced and direct $CP$ asymmetries, respectively.
 
In addition to these, the $B^0 \rightarrow K^0_S K^0_S K^0_S$ mode is 
also discussed. 
Due to Bose statistics, this final state is purely $CP$-even.
The BaBar collaboration has made an attempt to resolve intermediate states 
using Dalitz distribution.

\section{Time-dependent $B^0 \rightarrow K^+K^-K^0_S$ 
Dalitz analysis}

The time-dependent Dalitz analysis for $B^0 \rightarrow K^+K^- K^0_S$ has been 
performed by both the BaBar \cite{KKKS_babar} and Belle \cite{KKKS_belle}
Collaborations. Here, $\phi K^0_S$, $f_0 K^0_S$,
$f_X(1500) K^0_S$, $\chi_{c0} K^0_S$ and non-resonant contributions 
are taken into account in the fit to the time-dependent Dalitz distribution.
With the currently available statistics, we find multiple solutions.
In the Belle measurement, there are four solutions. The preferred solution can 
not be selected by the fit likelihood alone. 
With external information related to $f_0(980)$ and $f_X$ branching fractions
into charged pion and Kaon pairs, one of them is preferred, where 
$f_X$ is assumed as $f_0(1500)$.
In the BaBar measurement, the selected candidate events
are categorized into two groups, low and high $K^+K^-$ mass regions.
The most important process, $\phi K^0_S$ is contained in the low $K^+K^-$
mass region, and two solutions are found in the low mass region fit. 
Since those two solutions exhibited $\Delta \log({\cal L})=0.1$, 
the solution (1) in Ref. \cite{KKKS_babar} is presented as the 
preferred solution.

\begin{figure}
\centering
\includegraphics[width=0.4\textwidth]{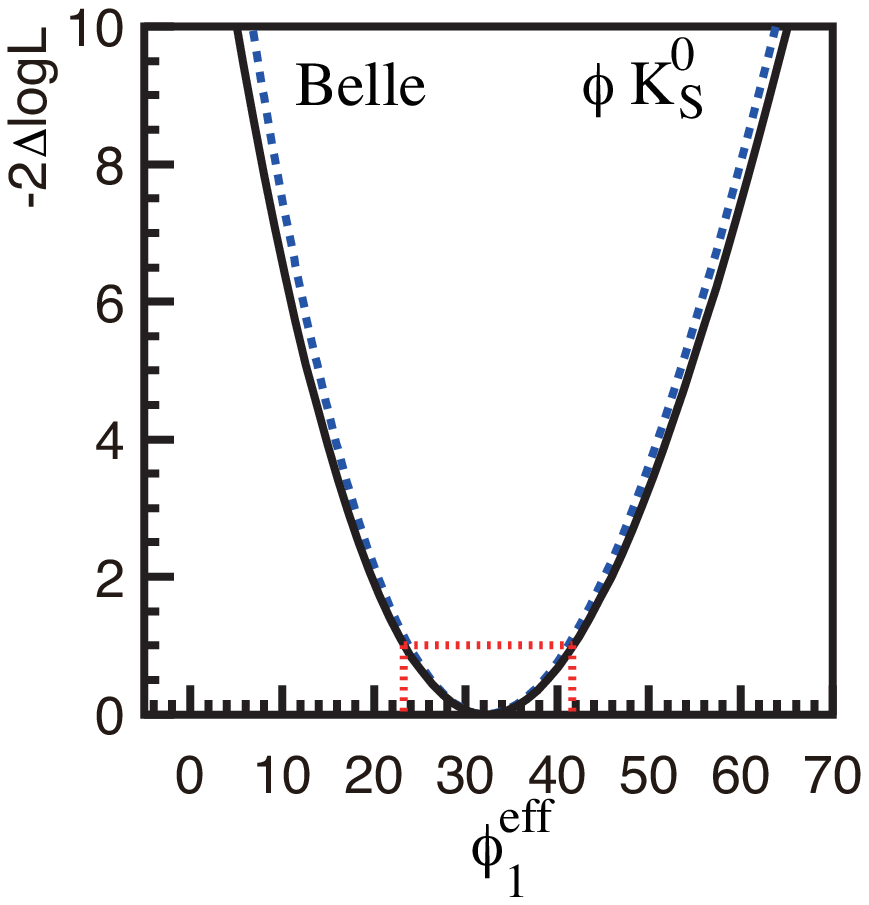}
\includegraphics[width=0.4\textwidth]{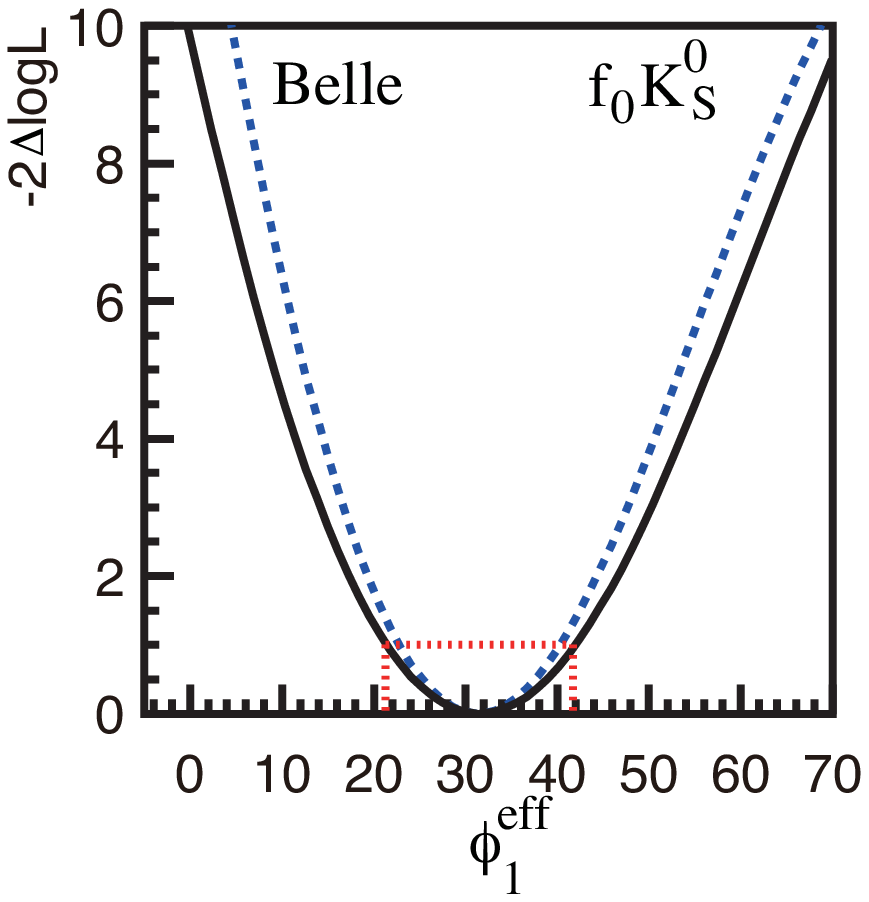}
\caption{Likelihood scans of $\phi_1^{\rm eff}$ (in degree) of
$B^0 \rightarrow \phi K^0_S$ (left) and 
$B^0 \rightarrow f_0 K^0_S$ (right) for the preferred solution
in the time-dependent Dalitz analysis 
for $B^0 \rightarrow K^+ K^- K^0_S$ decays at Belle.
The solid (dashed) curve contains the total (statistical) error 
and the dotted box indicates the parameter range corresponding 
to $\pm 1 \sigma$.}
\label{lhscan_phiks_belle}
\end{figure}

\begin{figure}
\centering
\includegraphics[width=0.4\textwidth]{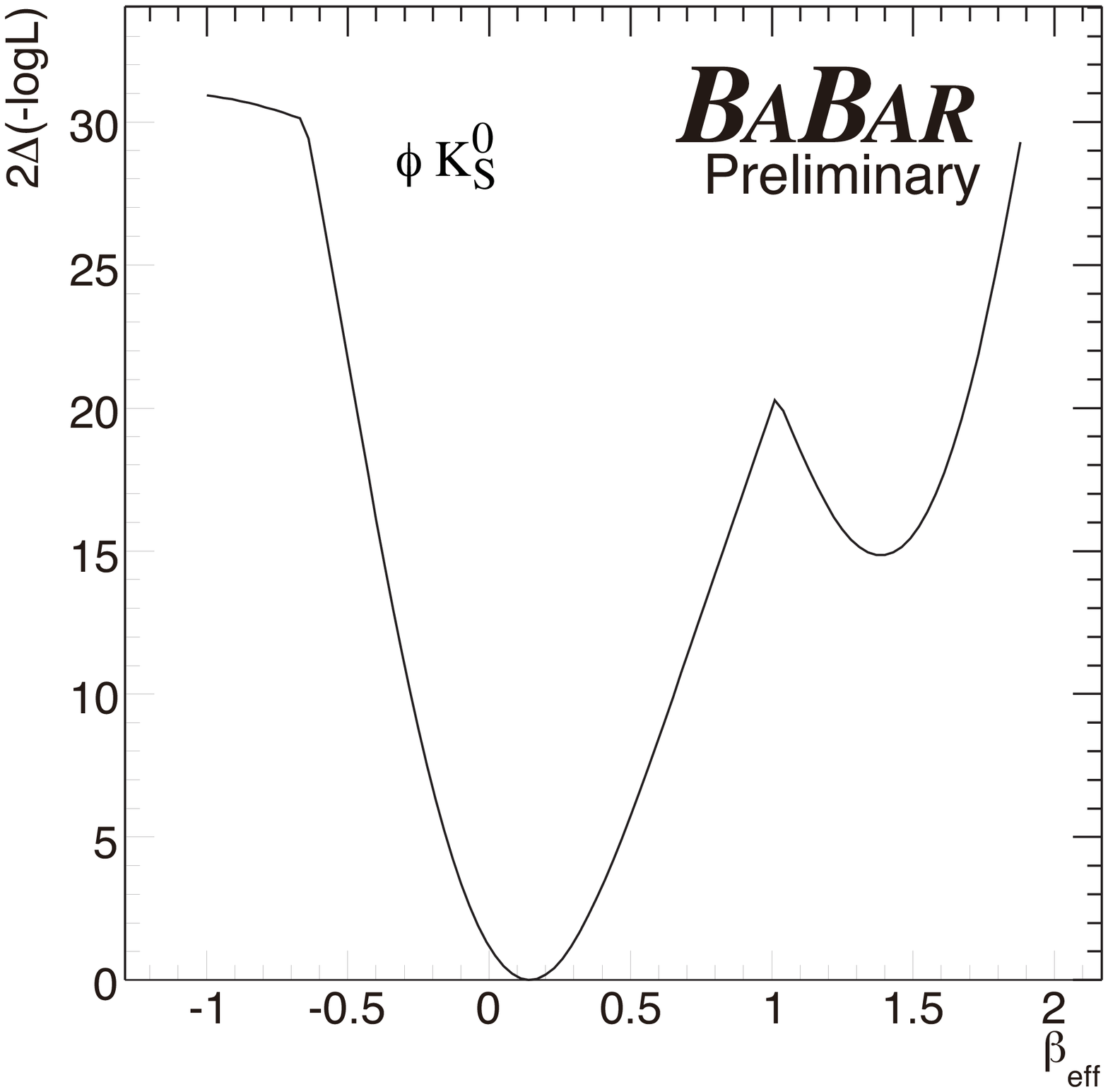}
\includegraphics[width=0.4\textwidth]{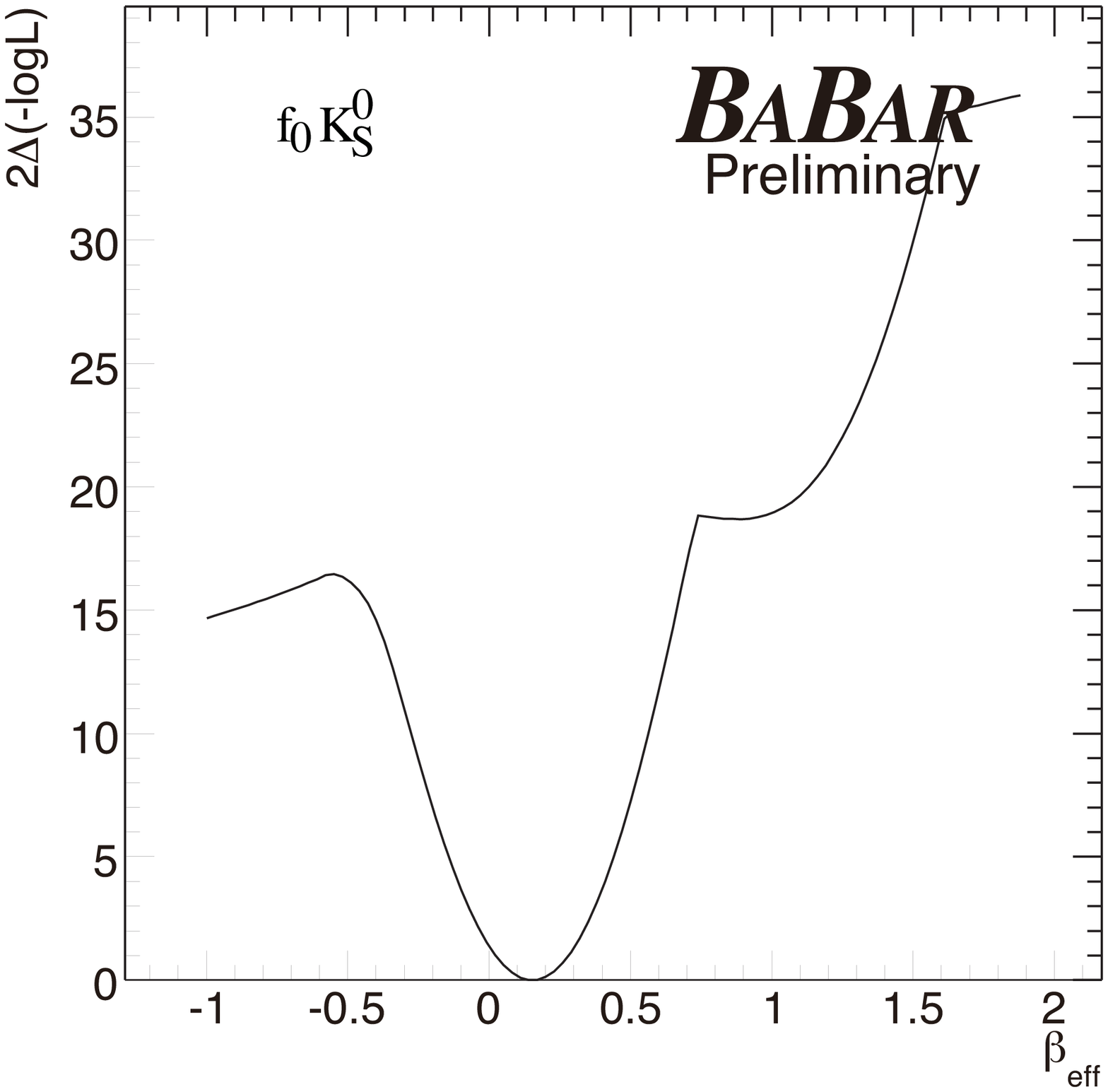}
\caption{Likelihood scans of $\phi_1^{\rm eff} = \beta_{\rm eff}$ 
(in radian) of $B^0 \rightarrow \phi K^0_S$ (left) and 
$B^0 \rightarrow f_0 K^0_S$ (right) for the preferred solution
of low mass fit for $B^0 \rightarrow K^+ K^- K^0_S$ decays at BaBar.}
\label{lhscan_phiks_babar}
\end{figure}

\begin{figure}[htb]
\centering
\includegraphics[width=0.4\textwidth]{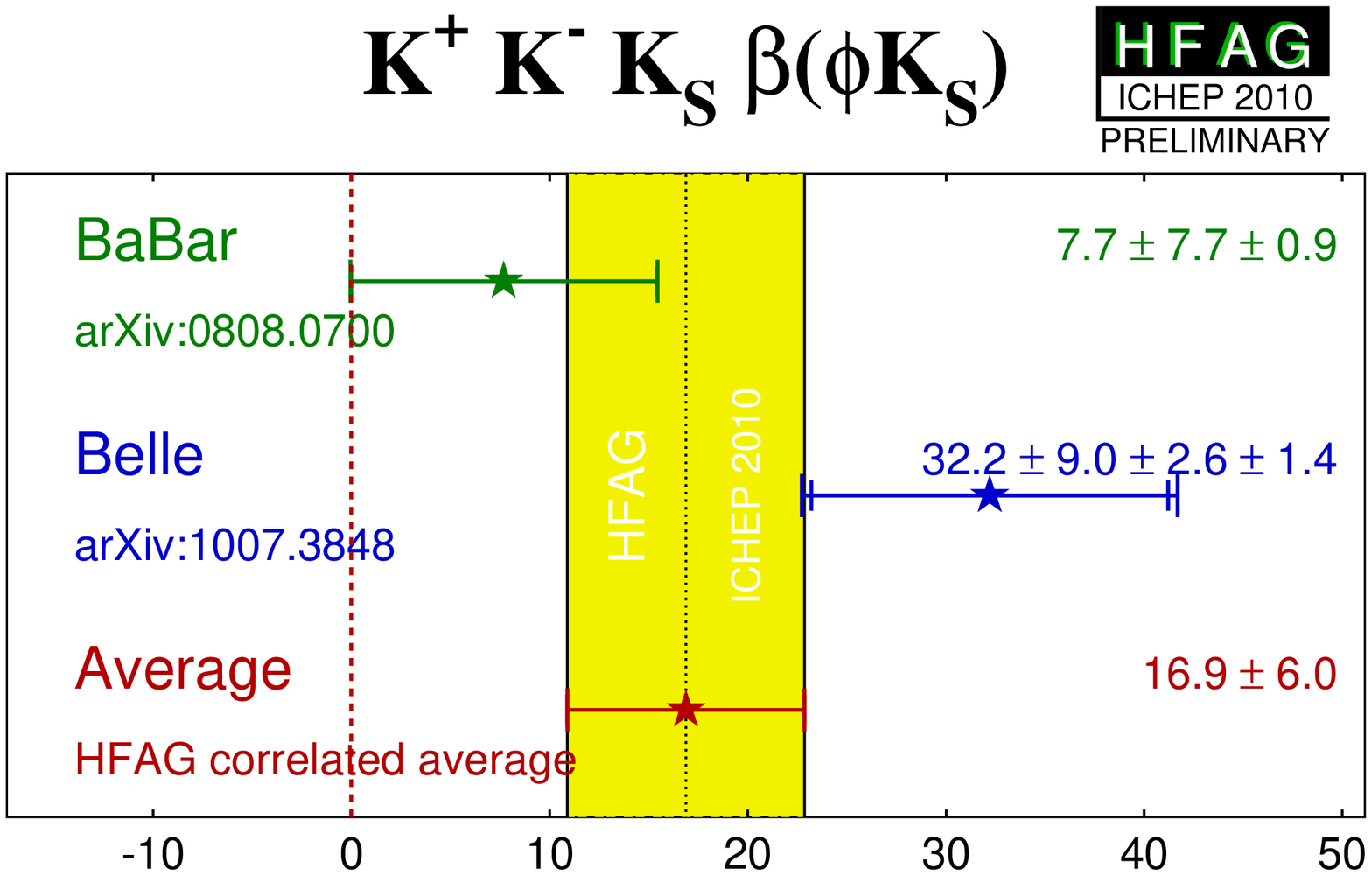}
\includegraphics[width=0.4\textwidth]{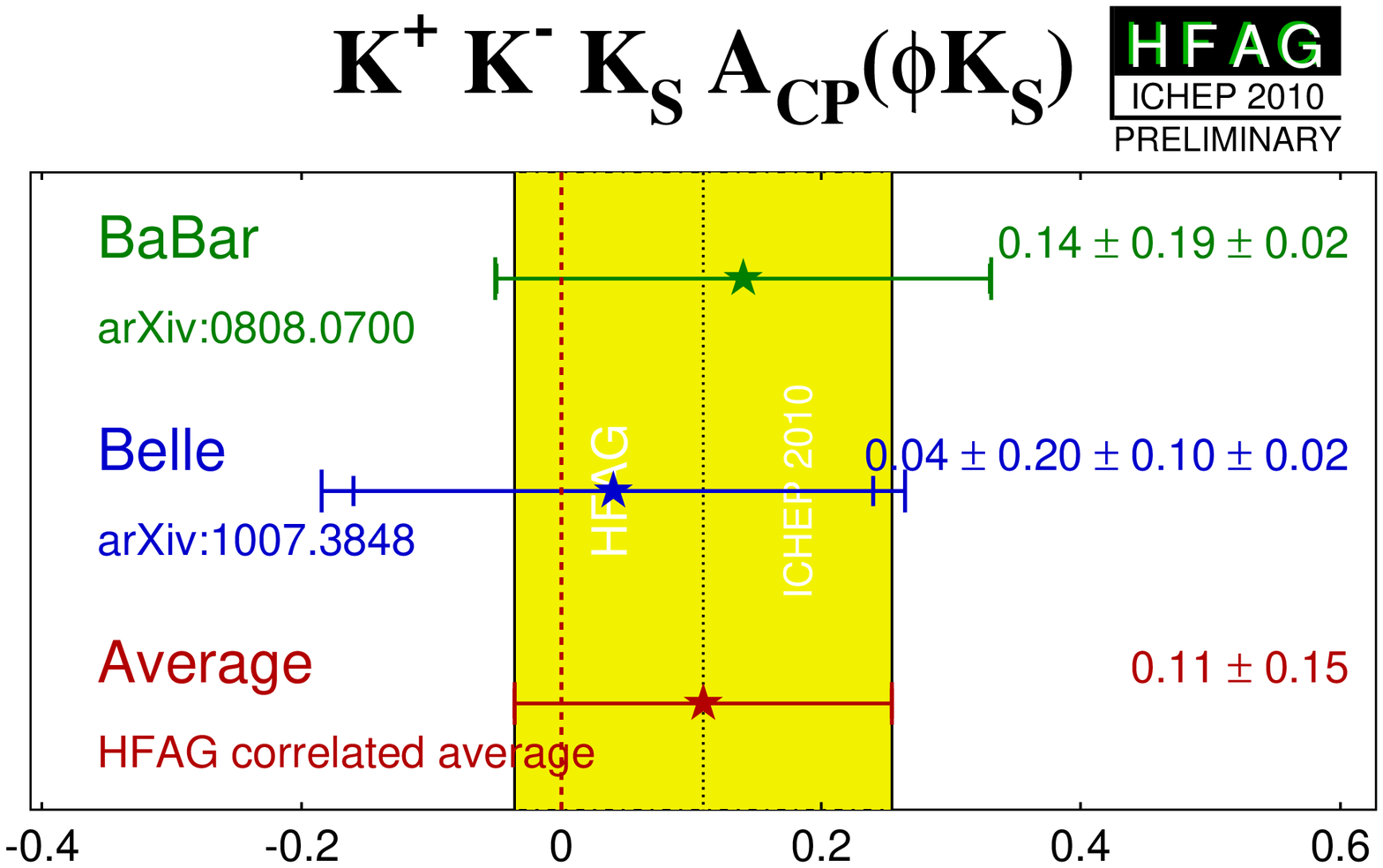}
\caption{The preferred solution's $CP$ violation parameters for 
$B^0 \rightarrow \phi K^0_S$.}
\label{fig_phiks}
\end{figure}

\begin{figure}[htb]
\centering
\includegraphics[width=0.4\textwidth]{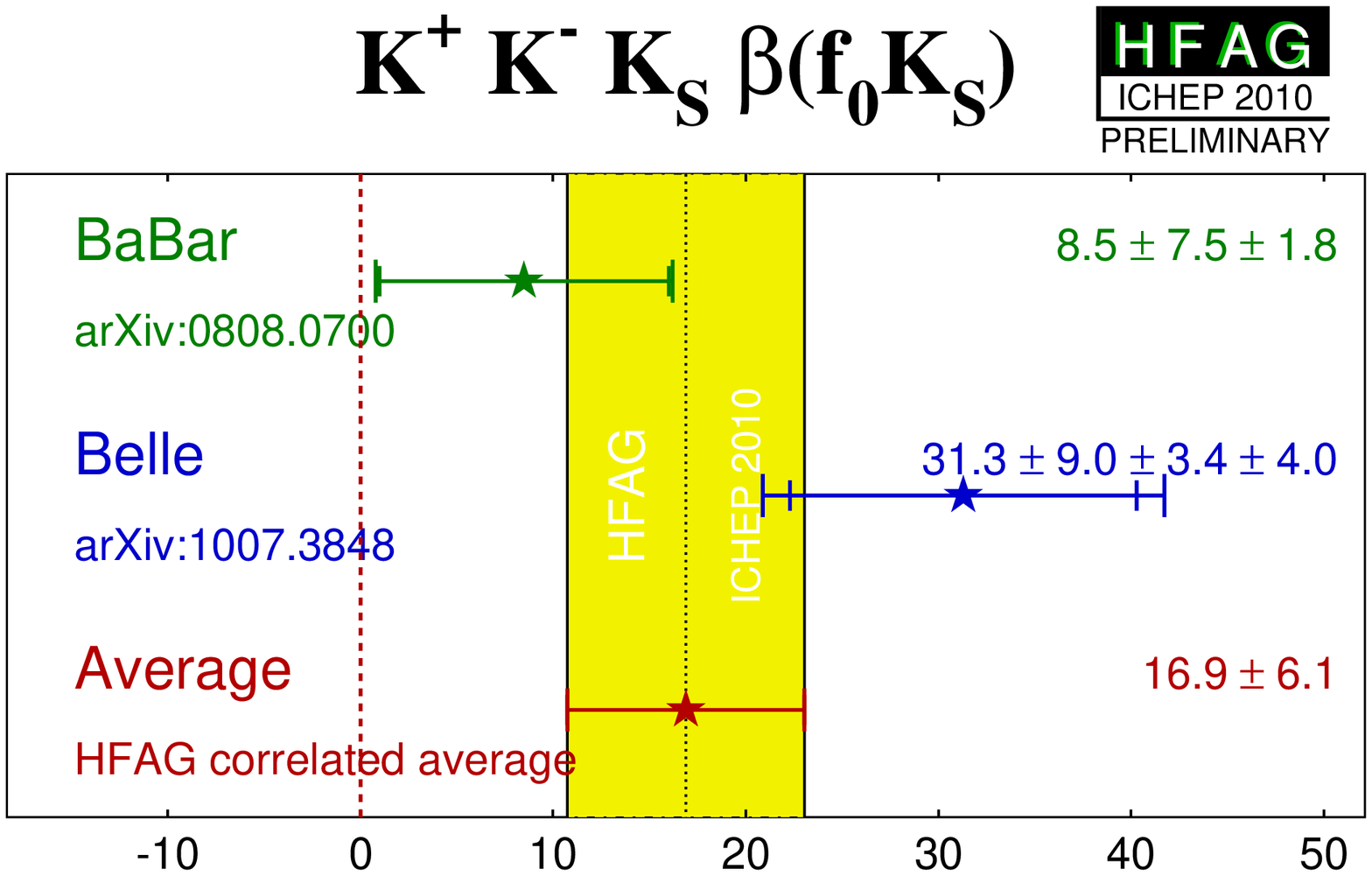}
\includegraphics[width=0.4\textwidth]{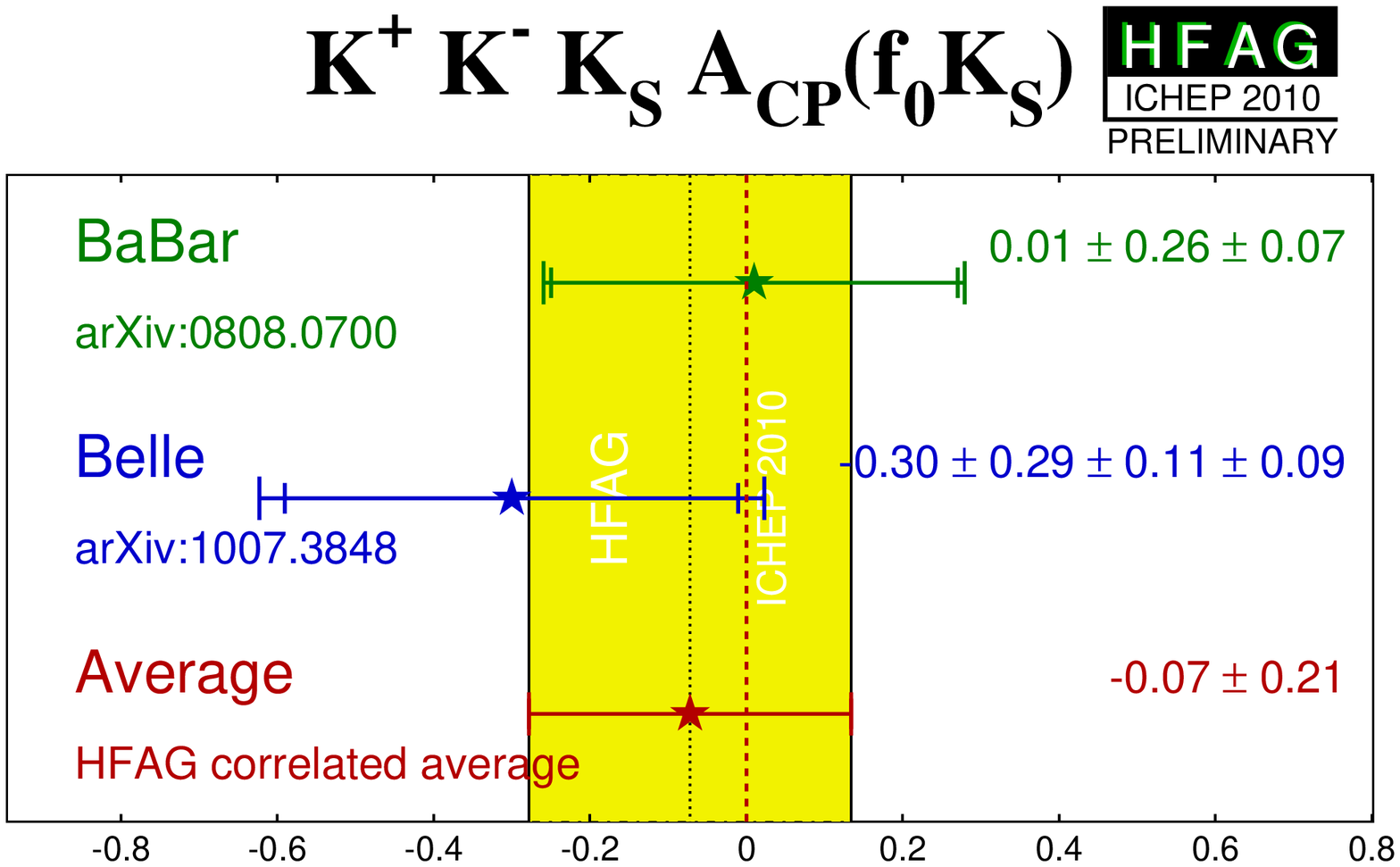}
\caption{The preferred solution's $CP$ violation parameters for 
$B^0 \rightarrow f_0 K^0_S$ in $f_0 \rightarrow K^+K^-$ mode.}
\label{fig_f0kkks}
\end{figure}

With higher statistics the maximum of the likelihood itself will 
determine the most preferred solution clearly.
The preferred solution results of each experiment are summarized in 
Figs. \ref{fig_phiks} and \ref{fig_f0kkks} for $B^0 \rightarrow \phi K^0_S$
and $B^0 \rightarrow f_0 K^0_S$, respectively.
So far, we have not observed significant deviation from the measurements 
with $B^0 \rightarrow (c\bar{c}) K^0$.


\section{Time-dependent $B^0 \rightarrow \pi^+\pi^-K^0_S$
Dalitz analysis}

\begin{figure}[htb]
\centering
\includegraphics[width=0.4\textwidth]{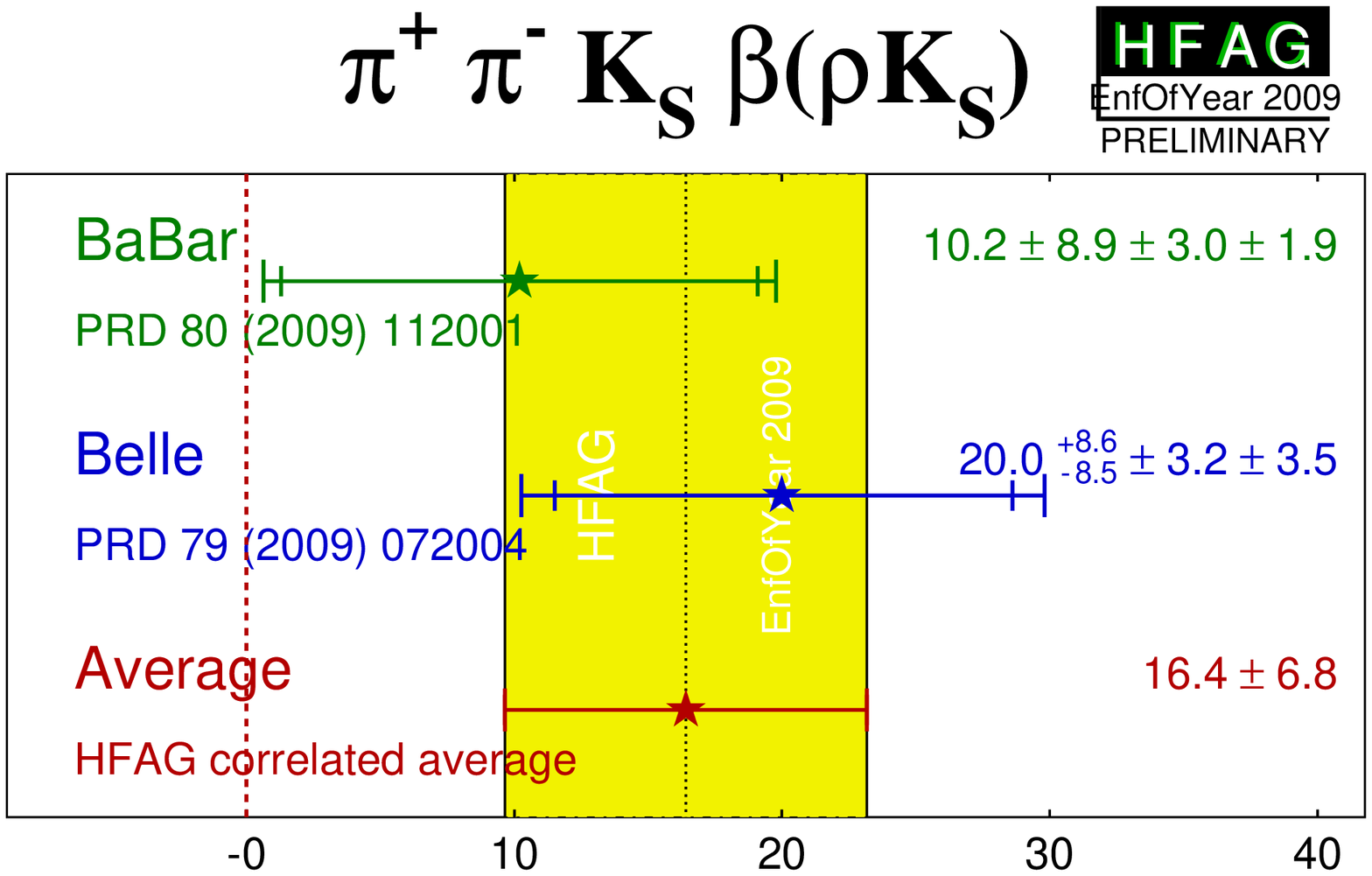}
\includegraphics[width=0.4\textwidth]{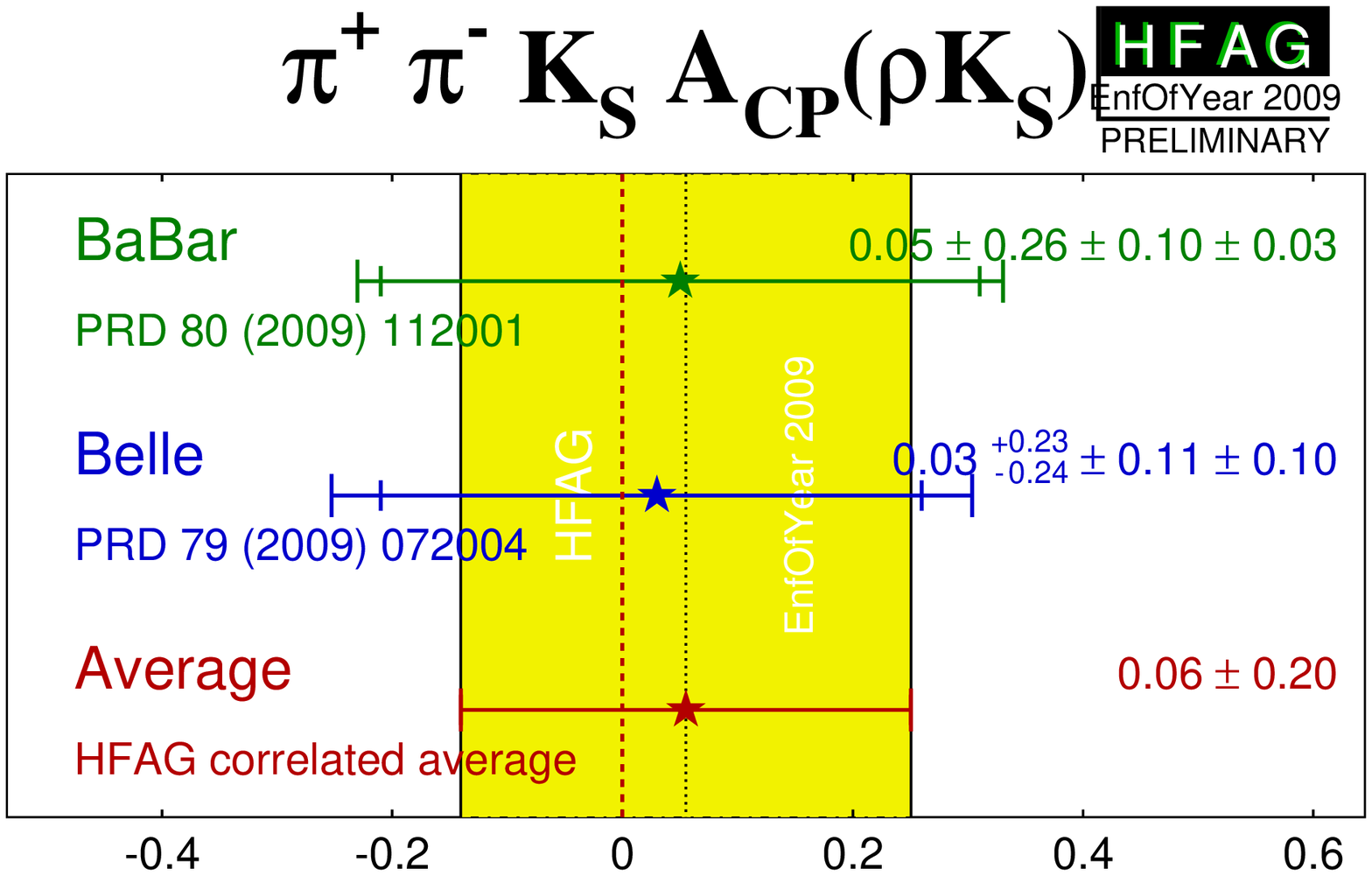}
\caption{The preferred solution's $CP$ violation parameters for 
$B^0 \rightarrow \rho^0 K^0_S$.}
\label{fig_rho0ks}
\end{figure}

\begin{figure}[htb]
\centering
\includegraphics[width=0.4\textwidth]{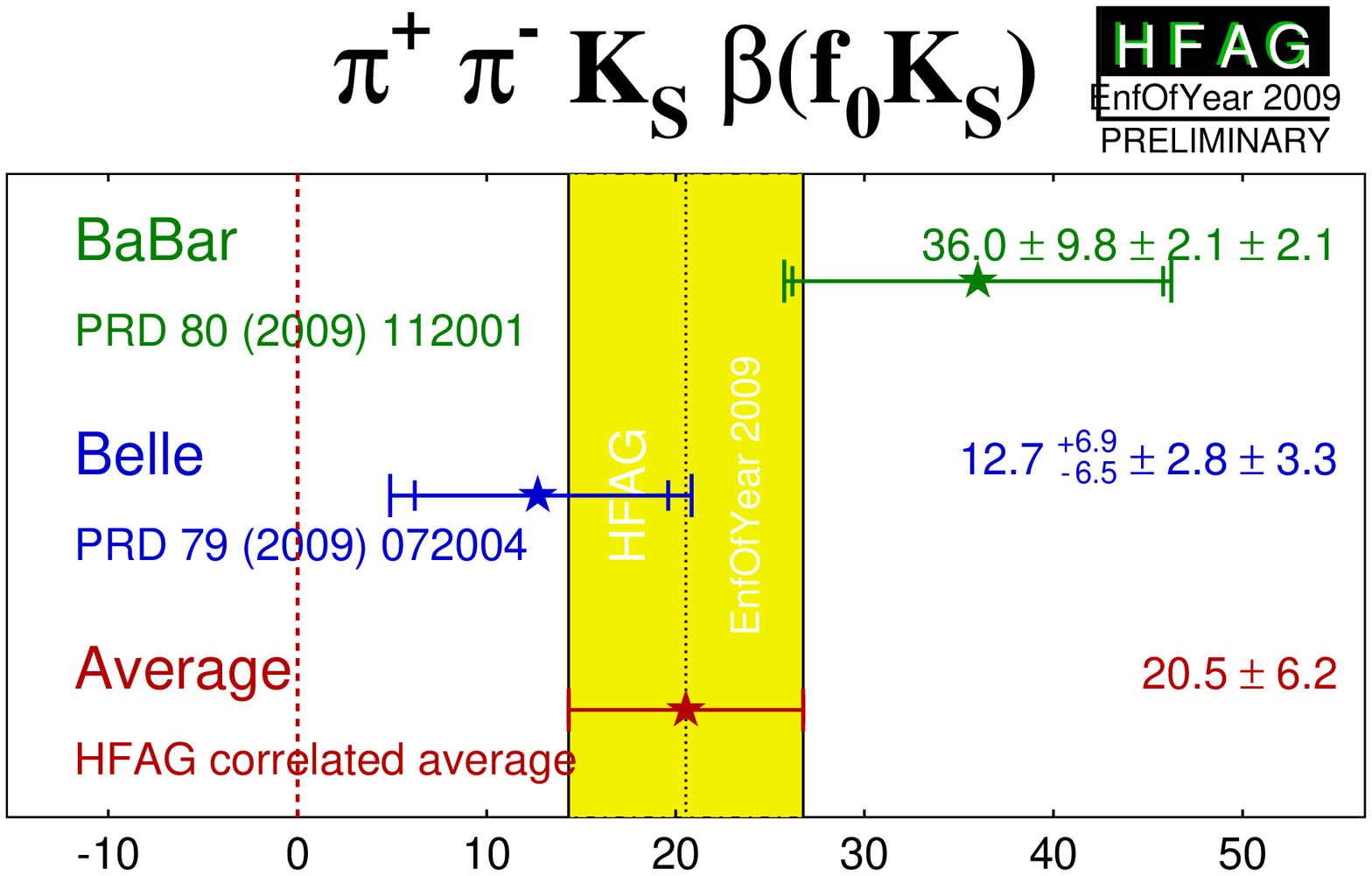}
\includegraphics[width=0.4\textwidth]{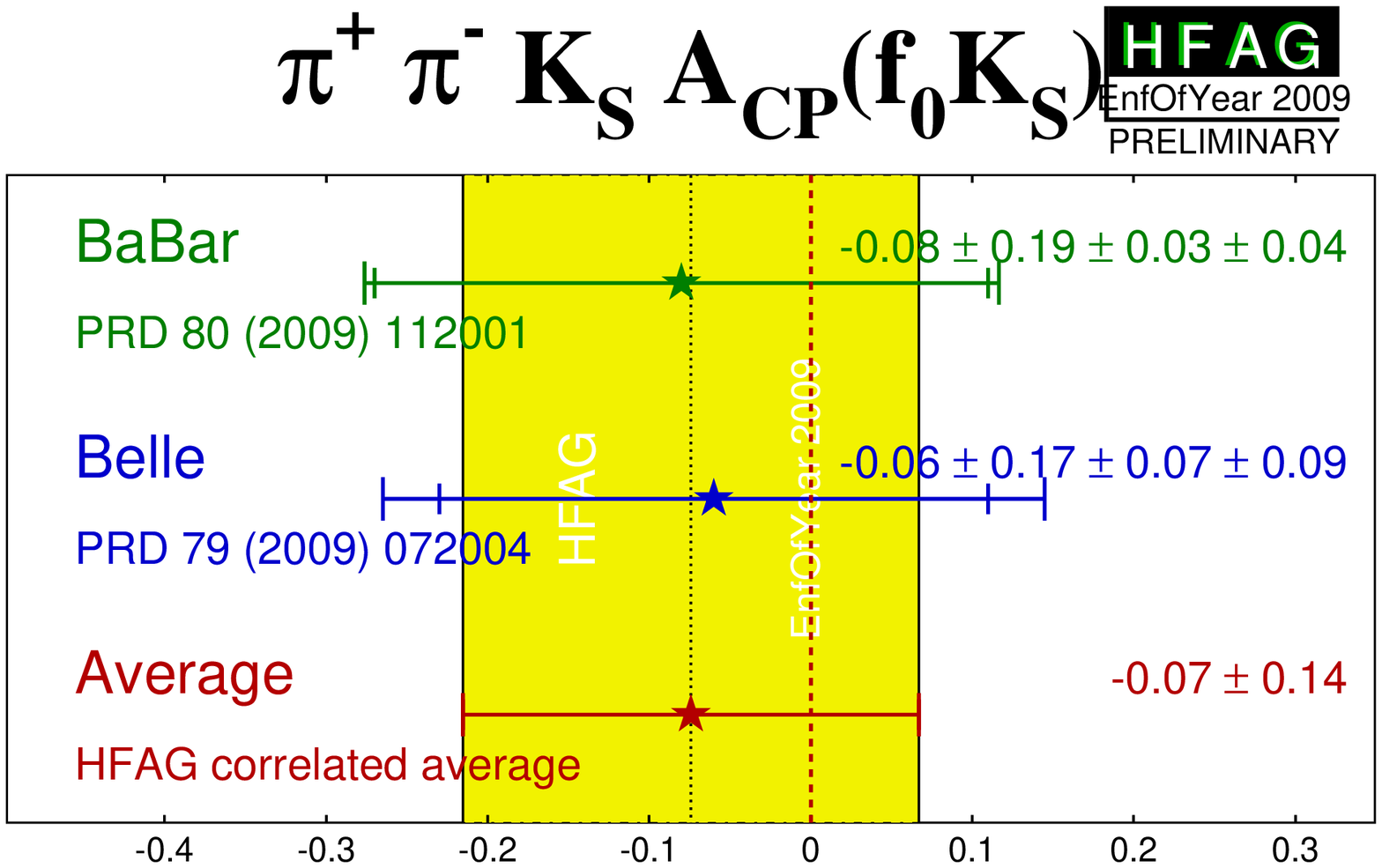}
\caption{The preferred solution's $CP$ violation parameters for 
$B^0 \rightarrow f_0 K^0_S$ in $f_0 \rightarrow \pi^+\pi^-$ mode.}
\label{fig_f0pipiks}
\end{figure}

In the $B^0 \rightarrow \pi^+\pi^- K^0_S$ decay, $\rho^0 K^0_S$ and 
$f_0(980) K^0_S$ are the main contributions. 
Since $\rho^0$, $f_0$ and other possible intermediate states have 
wide natural widths, they are overlapping in $\pi^+\pi^-$ invariant
mass distribution. 
Therefore a time-dependent Dalitz analysis is very effective to
extract $CP$ violation parameters while resolving interference among 
possible contributions.
The BaBar and Belle Collaborations have both performed this measurement
\cite{pipiKS_babar} \cite{pipiKS_belle}.
Similarly as the $B^0 \rightarrow K^+K^-K^0_S$ case, multiple solutions 
are found. BaBar and Belle found two and four solutions, respectively.
The solution presented as nominal is selected by an ensemble test as well as 
external information at Belle, while likelihood scan results are used 
to specify the most preferred solution at BaBar.
The results of the preferred solution of each experiment are summarized in 
Figs. \ref{fig_rho0ks} and \ref{fig_f0pipiks} for 
$B^0 \rightarrow \rho^0 K^0_S$
and $B^0 \rightarrow f_0 K^0_S$, respectively.
So far, no significant deviation from the measurements 
with $B^0 \rightarrow (c\bar{c}) K^0$ has been observed.


\section{$B^0 \rightarrow K^0_S K^0_S K^0_S$ mode}

This is a purely $CP$-even final state though there can be several
intermediate states \cite{b_to_ppp}. 
Since there are no up type quarks in the decay 
amplitude, it has very small Cabibbo-suppressed tree contribution and 
is thus theoretically clean. 
Because of this fact, time-dependent $CP$ violation measurements 
have been performed on the inclusive three-body final state.
The BaBar result \cite{ksksks_babar} is
\begin{eqnarray}
\sin 2 \phi_1^{\rm eff} &=& 0.90 ^{+0.18}_{-0.20} \mbox{(stat.)} 
^{+0.03}_{-0.04} \mbox{(syst.)} \\
{\cal A} = -{\cal C} &=& +0.16 \pm 0.17 \mbox{(stat.)} 
\pm 0.03 \mbox{(syst.)},
\end{eqnarray}
while the Belle measurement \cite{ksksks_belle} is
\begin{eqnarray}
\sin 2 \phi_1^{\rm eff} &=& 0.30 \pm 0.32 \mbox{(stat.)} 
\pm 0.08 \mbox{(syst.)} \\
{\cal A} = -{\cal C} &=& +0.31 \pm 0.20 \mbox{(stat.)} 
\pm 0.07 \mbox{(syst.)}.
\end{eqnarray}
In both experiments, precision is still limited by statistics
and therefore Super $B$-factory statistics is necessary 
to probe the New Physics effect as 
possible deviation from the values of $CP$ violation parameters
obtained by $B^0 \rightarrow J/\psi K^0$.

Recently, an attempt to resolve intermediate states has been 
carried out by the BaBar collaboration, where 200$\pm$15 events 
are found. A baseline model consists of $f_0 K^0_S$, $\chi_{c0} K^0_S$
and non-resonant contribution. Then a resonance is added and the 
likelihood is scanned varying its mass and width.  
As a result, $f_0(1710)$ and $f_2(2010)$ are found to be 
significantly contributing, while there was no evidence of 
$f_X(1500)$ \cite{ksksks_dalitz}.

\section{Summary}

In summary, for $B^0 \rightarrow K^+ K^- K^0_S$ and 
$B^0 \rightarrow \pi^+\pi^- K^0_S$ decays,
the time-dependent Dalitz analysis technique has been 
performed by both BaBar and Belle collaborations. 
In both experiments, multiple solutions are found and we are 
not always able to select one of them by the fit goodness alone.
Therefore the most preferred solution is identified with 
likelihood scan results, an ensembel test as well as external information,
depending on each analysis.
The preferred solutions' $CP$ violation parameters, 
$\phi_1^{\rm eff}$ and $\cal{A}_{CP}$ do not exhibit 
significant deviation from the ones obtained by 
$B^0 \rightarrow (c\bar{c})K^0$.
With Super $B$-factory level statistics, the best of the 
multiple solutions can be identified from likelihood alone.
In $B^0 \rightarrow K^0_S K^0_S K^0_S$ mode, 
$f_0(1710)$ and $f_2(2010)$ are found to give significant contributions 
in addition to the known components, $f_0 K^0_S$ and $\chi_{c0} K^0_S$.



\Acknowledgements
Author acknowledges support from MEXT, JSPS and Nagoya's TLPRC (Japan);
ARC and DIISR (Australia); NSFC (China); MSMT (Czechia);
DST (India); MEST, NRF, NSDC of KISTI (Korea); MNiSW (Poland); 
MES and RFAAE (Russia); ARRS (Slovenia); SNSF (Switzerland); 
NSC and MOE (Taiwan); and DOE (USA).

\end{document}